\documentclass[12pt]{article}

\newenvironment{sciabstract}{%
\begin{quote} \bf}
{\end{quote}}

\newcounter{lastnote}
\newenvironment{scilastnote}{%
\setcounter{lastnote}{\value{enumiv}}%
\addtocounter{lastnote}{+1}%
\begin{list}%
{\arabic{lastnote}.}
{\setlength{\leftmargin}{.22in}}
{\setlength{\labelsep}{.5em}}}
{\end{list}}

\usepackage{graphicx,amsmath,amssymb,bm}

\usepackage{color}

\title{Wave turbulence revisited: Where does\\ the energy flow?} 

\author
{ L.~V.~Abdurakhimov,$^{1,2}$ I.~A.~Remizov,$^{1}$  A.~A.~Levchenko,$^{1}$\\
G.~V.~Kolmakov,$^{3}$ Yu.~V.~Lvov$^{4}$\\
\\
\normalsize{$^{1}$Institute of Solid State Physics RAS, Chernogolovka, }\\
\normalsize{Moscow region, 142432, Russia}\\
\normalsize{$^{2}$Okinawa Institute of Science and Technology, Okinawa,}\\
\normalsize{904-0495,  Japan (present address)}\\
\normalsize{$^{3}$Physics Department, New York City College of Technology, }\\
\normalsize{City University of New York, Brooklyn, NY 11201, USA}\\
\normalsize{$^{4}$Department of Mathematical Sciences, Rensselaer Polytechnic Institute,}\\
\normalsize{Troy, NY 12180, USA}\\
}

\date{}

\begin{document}

\baselineskip24pt

\maketitle 

%
%
%
%
%
%
%
%
%
%

\begin{sciabstract}
  Turbulence in a system of nonlinearly interacting waves is referred to
as wave turbulence \cite{Zakharov:92}. It has been known since seminal
work by Kolmogorov \cite{Kolmogorov:41}, that turbulent dynamics is
controlled by a directional energy flux through the wavelength scales.
We demonstrate that an energy cascade in wave turbulence can be {\it
  bi-directional}, that is, can simultaneously flow towards large and
small wavelength scales from the pumping scales at which it is
injected.  This observation is in sharp contrast to existing
experiments and wave turbulence theory where the energy flux only
flows in one direction.  Establishment of the bi-directional energy
cascade changes the energy budget in the system and leads to formation
of large-scale, large-amplitude waves similar to oceanic rogue
waves \cite{Dean:90}. To study surface wave turbulence, we took
advantage of capillary waves on a free, weakly charged surface of
superfluid helium He-II at temperature $ \sim 1.7$ K, which
are identical to those on a classical Newtonian fluid with extremely low viscosity.  
\end{sciabstract}


Wave turbulence, or turbulence in a system of
interacting waves, is manifested in various physical systems including
atmospheric waves \cite{Smith:05}, the earth's magnetosphere and its
coupling with the solar wind \cite{Southwood:78}, shock propagation in
Saturn's bow,\cite{Scarf:81} interstellar plasmas \cite{Bisnovatyi:95},
and ocean wind-driven waves \cite{Toba:73}. Wave turbulence is much
easier to understand than hydrodynamic turbulence in an incompressible
liquid, because it is appropriate when the building blocks of a system
are linear waves that admit analytical descriptions.  Wave turbulence
theory \cite{Zakharov:92} based on a kinetic equation for a wave
ensemble predicts a steady-state scale-invariant solution that
describes a constant flux of energy towards smaller scales, which is
referred to as a direct energy cascade.  Such a power-law spectrum can
be viewed as the wave analog of the Kolmogorov spectrum of
hydrodynamic turbulence~\cite{Kolmogorov:41,Frisch:95} and is referred
to as the Kolmogorov-Zakharov (KZ) spectrum of wave
turbulence \cite{Zakharov:92}. Direct cascade of wave turbulence has
been extensively studied in experimental and theoretical
works \cite{Zakharov:67,Pushkarev:96,Wright:97,Henry:00,Abdurakhimov:09,Deike:14}.
In what follows, we focus on surface capillary waves on a fluid
surface, that is, short waves for which surface tension is the primary
restoring force.  The dispersion relation between the wave frequency
$\omega$ and the wave number $k$ for pure capillary waves is
\begin{equation}
\omega(k) = \sqrt{\alpha k^3 \over \rho}, \label{eq:w}
\end{equation}
where $\alpha$ is the surface tension and $\rho$ is the fluid density.
Capillary waves are important for the energy and momentum
transfer on a fluid surface \cite{Zakharov:92}, and for  the transfer of
gas into solution through a gas-liquid interface \cite{Szeri:97}.

In this report, based on the results of experimental and numerical
studies, we report that in sharp contrast to existing theory and
experiments, the energy flux of weakly nonlinear capillary waves can
also propagate towards the large-scale, low-frequency spectral region
{\it simultaneously} with a conventional direct cascade.  Formation of
this bi-directional turbulent cascade results in significant changes
in the energy budget of the system.  Specifically, small-scale
turbulent oscillations are suppressed, whereas sustained
high-amplitude large-scale oscillations are formed.  A bi-directional
cascade of energy was recently predicted for the two-component
hydrodynamics in the solar wind \cite{Che:14}. However, such a cascade
has never been observed or predicted for capillary waves.  Moreover,
it has never been observed for systems in which resonant three-wave
interactions dominate and no additional integrals of motion are
present. We demonstrate that it is the finite viscous damping in the
low-frequency domain that results in the bi-directional cascade
formation.

We study capillary waves on the surface of superfluid helium (He-II)
at temperature $T \sim 1.7$~K.  He-II demonstrates many quantum
features, among which are the famous fountain effect in response to
heating, extremely high heat conductivity, and quantization of
vorticity in the fluid bulk \cite{Khalatnikov:65}. Nevertheless,
oscillations of a free He-II surface behave much like surface
oscillations of a classical fluid with very low
viscosity \cite{Khalatnikov:65,Abdurakhimov:09,Abdurakhimov:10}. He-II
provides an ideal testbed for studying nonlinear wave dynamics due to
the possibility of driving the weakly-charged He-II surface directly
by an oscillating electric field, virtually excluding the excitation
of bulk modes \cite{Brazhnikov:02a}. This method is similar to the
oceanographic case where waves are generated due to wind drag applied
directly to the fluid surface.  
Previous experiments with waves on quantum fluids (liquid helium and
hydrogen) allowed detailed study of the direct cascade of capillary
turbulence \cite{Abdurakhimov:09}, including modification of the
turbulent spectrum by applied low-frequency
driving \cite{Brazhnikov:05}, and the turbulent bottleneck phenomena in
the high-frequency spectral domain \cite{Abdurakhimov:10}.
Generation of pure surface waves
without creating bulk vorticity can  hardly be achieved in experiments
with conventional fluids like water or mercury, in which the waves are
launched by applying vertical high-frequency oscillations to a
container \cite{Henry:00,Shats:10} or by moving flaps
immersed in the fluid \cite{Falcon:07}. Coupling of such surface waves with
bulk vorticity modifies the surface dynamics \cite{Savelsberg:08}.

In our experiments helium was condensed into a cylindrical cup formed
by a bottom capacitor plate and a guard ring, and was positioned in a
helium cryostat.  The free surface of the liquid was positively
charged as the result of $\beta$--particle emission from a radioactive
plate located in the bulk liquid. Oscillations of the liquid surface
were excited by application of an AC voltage $U(t) = U_d \sin(\omega_d
t)$ to the upper capacitor plate.  Oscillations of the fluid surface
elevation $\zeta(\bm{r},t)$ were detected through variations of the
power $P(t)$ of a laser beam reflected from the surface (Fig.\ 1A).
(Here, $t$ is time and $\bm{r}$ is the two-dimensional coordinate in
the surface plane). The capillary wave power spectrum $\zeta({\omega})
\propto P({\omega})$ was calculated via the Fourier time transform of
the signal $P(t)$ \cite{Brazhnikov:02a}. The finite size of the cell
results in a discrete wave number spectrum.  Figure 1B shows a
snapshot made through the cryostat glass of turbulent waves on the
helium surface. Large-scale waves with lengths much larger then those
at the driving frequency $\omega_d$ are clearly seen.

Figures 1C,D show the evolution of the ensemble-averaged turbulent
wave spectrum $I(\omega)= \langle |\xi(\omega)|^2 \rangle$ with
increasing driving amplitude $U_d$, when the driving frequency is
$\omega_d / 2 \pi = 68$ Hz.  In Fig.\ 1C for a moderate pumping
$U_d=4$~V, the direct Kolmogorov-Zakharov cascade  forms in the
high-frequency domain $2\times 10^2$ Hz $< \omega / 2 \pi < 2 \times
10^3$ Hz. At very high frequencies $\omega / 2 \pi \sim 2 \times 10^3$
Hz, the Kolmogorov-Zakharov cascade is terminated by bulk viscous
damping.  Weak low frequency oscillations at $\omega<\omega_d$, with
$I(\omega) \leq 10^{-11}$~cm$^2$s in Fig.~1C, are caused by mechanical
vibrations of the experimental setup.

With an increased driving voltage of $U_d = 14$~V in Fig.~1D, there
are many low-frequency peaks in the spectrum that have heights a few
orders of magnitude larger: $ I(\omega) \approx 10^{-7} -
10^{-6}$~cm$^2$s.  Calculations of the wave energy\cite{Brazhnikov:02a}
 \[
E = {\alpha } \int |\nabla \zeta(\bm{r},t)|^2 d\bm{r} 
\] 
from the data in Fig.~1D shows that only about $1$\% of the wave
energy is concentrated in the high-frequency domain $\omega \geq
\omega_d$, whereas 99\% of the energy is localized at frequencies
$\omega<\omega_d$.

To understand the formation of large-amplitude low-frequency waves, we
performed numerical modeling of the wave dynamics in the cylindrical
cell with external driving and viscous damping. In Fig.\ 2A (red
pulses) the steady-state wave spectrum $I({\omega})$ is similar to
that observed in the experiment for high-amplitude driving $U_d=14$ V
(cf.\ Fig.~1C).  In the domain $\omega > \omega_d$, the high-frequency
spectrum forms in agreement with current and previous observations.
We found it highly surprising that, in both the experiment and
simulations, the low-frequency waves with $\omega < \omega_d$ retain
finite values; moreover, the amplitudes of some low-frequency waves
exceeds those at the driving frequency $\omega_d$.

To explain the formation of the low-frequency waves, we demonstrate
that bi-directional energy flux is established in the system in place
of the traditional direct energy cascade.  In the simulations, we
varied the low-frequency damping and kept all other parameters fixed.
Low-frequency damping is the result of viscous drag at the cell
bottom \cite{Christiansen:95}, and high-frequency damping is caused by
bulk viscosity in the fluid \cite{Frisch:95}. We analyze the energy
balance in the system in the form of the continuity equation for
energy \cite{Frisch:95,Zakharov:92},
\begin{equation}
     {d E(\omega) \over dt} + {\Pi} = - \Gamma (\omega) + S(\omega), \label{eq:balance}
\end{equation}
where $E(\omega) = \int_0^{\omega} \varepsilon(\omega^\prime)
d\omega^\prime$ is the total wave energy in the spectral domain
$\omega^\prime < \omega$, $\varepsilon(\omega) = 2\pi \alpha
k(\omega)$ $\times(dk(\omega)/d\omega) \omega I(\omega)$ is the
spectral energy density, ${\Pi}$ is the total energy flux,
$\Gamma(\omega) = 2 \int_0^{\omega} \gamma(\omega^\prime)
\varepsilon(\omega^\prime) d\omega^\prime$ is the energy loss due to
viscous damping, and $S(\omega)$ is the energy source from the
driving. (Here $k(\omega)$ is found by inverting the dispersion
relation Eq.\ \ref{eq:w}.) In the steady state ${d E(\omega) / dt} = 0$,
the total energy balance in the low-frequency spectral domain $\omega<
\omega_d$ is
\begin{equation}
      {\Pi} =-  4 \pi  \alpha \int_0^{\omega}   \gamma(\omega^\prime)  k(\omega^\prime)  
      \left({dk(\omega^\prime)\over d\omega^\prime} \right) \omega^\prime I(\omega^\prime) d \omega^\prime,
      \label{eq:p}
\end{equation} 
because the source term is $S(\omega)$ is absent for low frequencies.
To investigate the dependence of the energy flux on the system
parameters, we calculated ${\Pi}$ from Eq.\ \ref{eq:p} for different
low-frequency damping coefficients and two cell radii (see
Fig.\ 2B). In the absence of low-frequency damping, the
thermodynamic-equilibrium Rayleigh-Jeans-like spectrum $I(\omega)
\propto \omega^{-1}$ is formed at $\omega < \omega_d$ (Fig.~2A, blue
squares).  This spectrum produces no energy flux through the frequency
scales \cite{Balkovsky:95}. The negative sign of ${\Pi}$ for finite
low-frequency dampings (Fig.\ 2B) corresponds to the flux direction
from the driving scales, $\omega \sim \omega_d$, towards the
low-frequency domain.

Wave turbulence predicts that the probability distribution function
(PDF) for wave amplitudes with specified wave numbers is a Gaussian
function. We verified numerically that this is indeed the case for
most modes. However, some modes showed significant deviations from the
predicted Gaussian form when low-frequency damping is applied, as
shown by the 10th mode in Fig.\ 2C.  The non-Gaussian tails in the PDF
in the presence of the bi-directional  energy cascade correspond to an
increased probability of the resonant formation of large-amplitude waves,
which may be thought as a capillary-wave analogue of ``rogue'' waves
observed in the ocean \cite{Dean:90}.

In conclusion, we demonstrated that energy flux from the driving scale
towards the damping region can be formed for capillary waves even if
the damping occurs at frequencies lower that the driving frequency.
This bi-directional energy flux provides a continuous energy source
for sustained low-frequency wave oscillations in the presence of
finite damping.  Furthermore, bi-directional energy flux provides
an effective global coupling mechanism between the scales.
In our experiments, we studied nonlinear capillary waves on the
surface of superfluid He-II. However, the concept of bi-directional energy
flux is relevant for a wider range of 
nonlinear systems, such as waves on classical fluids in wave
tanks \cite{Lukaschuk:09} and in restricted geometries
\cite{Herbert:10}, vibrating elastic plates \cite{Miquel:13}, 
and in quantum fluids \cite{Ganshin:08}. 
\noindent \vspace{0.1cm}

\bibliography{lowtempgk}
\bibliographystyle{Science}

\begin{scilastnote}
\item The authors are grateful to Prof.\  Leonid P.\ Mezhov-Deglin and
Prof.\ William L. Siegmann for valuable discussions.  G.V.K. gratefully
acknowledges support from the Professional Staff Congress -  City University of New York award 
\#66140-00 44.  Yu.V.L. is grateful for support to ONR, award
\#N000141210280. L.V.A, A.A.L and I.A.R. are grateful to the Russian
Foundation for Basic Research for support, grant \#13-02-00329. 
The authors gratefully acknowledge the Center for Theoretical Physics at
New York City College of Technology of the City University of New York for providing 
computational resources. 

A.A.L., G.V.K. and Yu.V.L. designed the research; A.A.L., L.V.A. and
I.A.R. performed the experiments; G.V.K. and Yu.V.L. developed the
model and performed the simulations; L.V.A, A.A.L, I.A.R., G.V.K, and
Yu.V.L. analyzed the data; and Yu.V.L., G.V.K.  and A.A.L. wrote the
paper.
\end{scilastnote}


\clearpage
 
\begin{figure*}[t]\label{fig:ONE}
\begin{center}\includegraphics[width=15.cm]{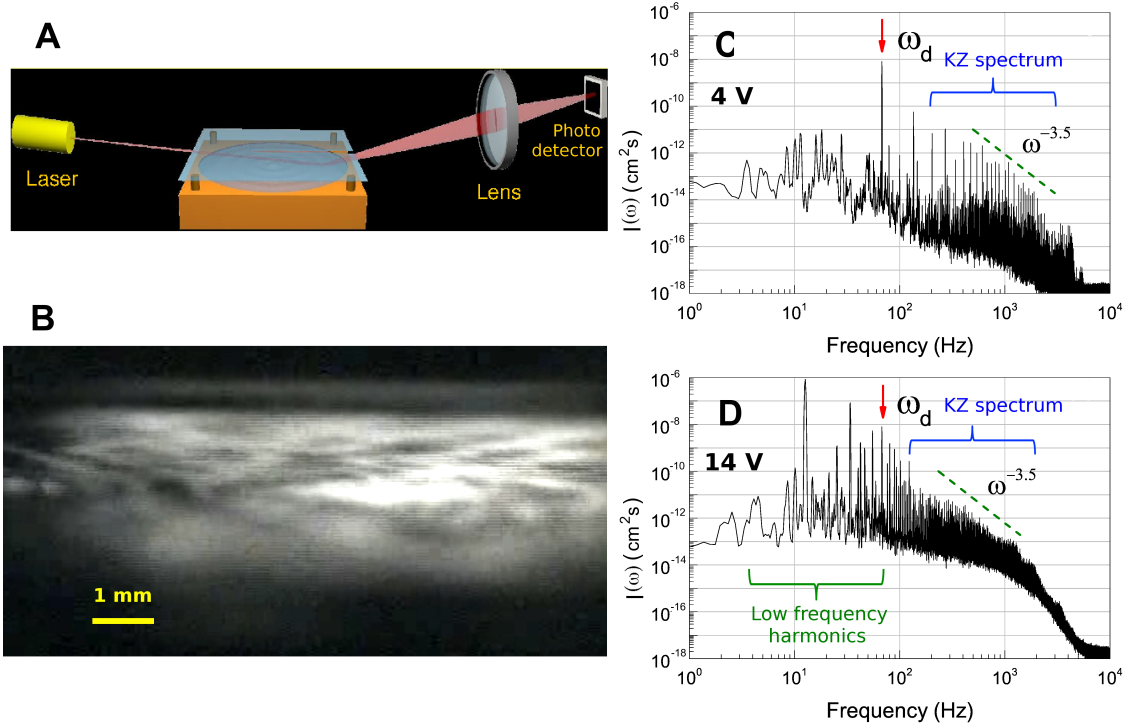}
\end{center} \vspace{-0.7cm}
 \caption{({\bf A}) Schematic of the experimental setup.
 Oscillations of the liquid helium surface are detected via variations
 of the total power of the reflected laser beam.  An optical cryostat
 containing the cell is not shown.  ({\bf B}) Snapshot of the turbulent
 surface of superfluid helium taken through the cryostat windows in
 the reflected light.  Large scale waves are clearly visible.  The
 driving frequency is $\omega_d / 2 \pi = 113$ Hz. A horizontal bar
 shows the length scale.  ({\bf C,D}) Formation of large-amplitude
 waves on the surface of superfluid helium at $\omega < \omega_d$ by
 increasing the AC driving voltage from $U_d=4$~V ({\bf c}) to 14 V
 ({\bf B}).  The driving frequency (arrow) is $\omega_d/2\pi = 68$ Hz.
 The wavelength at the driving frequency $\omega_d$ is $\simeq 780$
 $\mu$m. The conventional direct Kolmogorov-Zakharov (KZ) spectrum of
 capillary turbulence $I({\omega}) \propto \omega^{-3.5}$ between
 $2\times 10^2$ Hz $< \omega / 2 \pi < 2 \times 10^3$ Hz is shown by
 the dashed line in plates {\bf C} and {\bf D}.  Formation of
 low-frequency harmonics at $\omega < \omega_d$ with amplitudes larger
 than those at the driving frequency $\omega_d$, in addition to KZ
 spectrum, are clearly visible for high-amplitude driving $U_d=14$ V
 in {\bf D}.}
\end{figure*}
\clearpage

\begin{figure*}[t]\label{fig:TWO}
\begin{center}\includegraphics[width=15.cm]{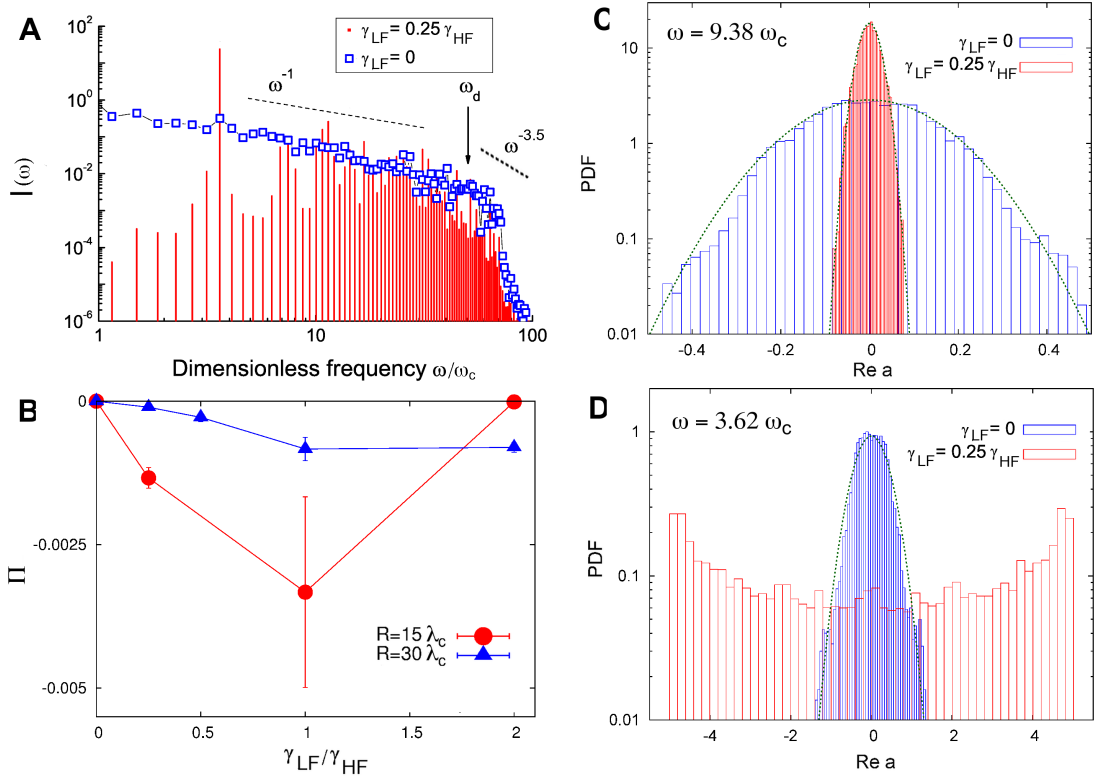}
\end{center} \vspace{-0.7cm}
\caption{({\bf A}) Numerical steady-state spectrum $I(\omega)$
 of sustained surface oscillations in the presence of low- and
 high-frequency damping (red peaks) and in the presence of
 only high-frequency damping (blue open squares). The spectrum is shown in units of 
 $\lambda_c^2 t_c$. The surface is driven at a
 frequency $\omega_d$ of the 50th resonance (arrowed).  The
 high-frequency Kolmogorov-Zakharov spectrum $I(\omega) \propto
 \omega^{-3.5}$ is formed for $\omega > \omega_d$, as consistent with
 Fig.\ 1D.  With the absence of low-frequency damping, $\gamma_{LF} =
 0$, the numerical wave spectrum (blue open squares) approaches the
 thermal-equilibrium spectrum $I({\omega}) \propto \omega ^ {-1}$ that
 carries no energy flux. The power-law spectral behavior $I(\omega)
 \propto \omega^{-3.5}$ and $\omega^{-1}$ are shown by dashed lines.
 The curve connecting the squares is shown to guide the eye.  The
 radius of the cylindrical cell is $R=15\lambda_c$; $\omega_c = (\rho
 g^{3} / \alpha)^{1/4}$ and $\lambda_c = (\alpha /\rho g)^{1/2}$ are
 used as units of frequency and length, respectively.  ({\bf B}) Energy
 flux ${\Pi}$ in units of $\alpha \omega_c$, incoming to the spectral
 domain $\omega < \omega_d$, that supports sustained low-frequency
 oscillations of the fluid surface, as a function of the low-frequency
 damping coefficient $\gamma_{LF}$.  The simulations were performed
 for the cell radii $R=15 \lambda_c$ (red circles) and $30 \lambda_c$
 (blue triangles). Vertical bars show the fluctuations of the flux as
 a standard deviation about the mean.  The line segments connecting
 the points are shown to guide the eye.  ({\bf C,D}) Probability
 distribution functions of $Re(a)$ for the 20th resonant mode at
 frequency $\omega = 9.38 \omega_c$ ({\bf C}) and for the 10th mode at
 frequency $\omega =3.62 \omega_c$ ({\bf D}) in the absence of
 low-frequency damping, $\gamma_{LF}=0$ (blue), and at
 $\gamma_{LF}=0.25 \gamma_{HF}$ (red).  PDFs are calculated for the
 spectra shown in {\bf A}.  Green lines show the Gaussian fit to the
 PDFs.  It is seen in {\bf D} that the Gaussian function fits the PDF
 for the 10th mode only if there is no low-frequency damping (blue
 bars). When the low-frequency dumping is applied and the
 bi-directional energy flux is thereby established, the PDF for the
 10th mode (red bars in {\bf D}) is far from Gaussian.}
\end{figure*}

\clearpage

\centerline{\bf Supplementary Materials}\vspace{0.1cm}

\noindent {\bf 1. Experimental techniques.}  The experimental
arrangements were similar to those in our previous experiments with
superfluid helium and liquid hydrogen
\cite{Abdurakhimov:09}. The cup in which the helium was
condensed has inner radius $R=30$ mm and depth $4$ mm.  The
experiments were conducted at temperature $T=1.7$ K of the superfluid
liquid.  The power $P(t)$ was measured with a photodetector and
sampled with an analog-to-digital converter.  The
capillary-to-gravity wave transition on the surface of superfluid
helium is at frequency $\sim 25$ Hz.  The finite depth of the waves
only influences the wave dispersion $\omega=\omega({k})$ at low
frequencies $\omega < 10$ Hz. The capillary wave length for superfluid helium 
is $\lambda_c = 0.17$ cm. 
%
\vspace{0.2cm}
The kinematic viscosity of He-II at $T=1.7K$ is $\nu = 2.6 \times
10^{-4}$ cm$^2$/s, which  is $\sim 40$ times lower than that for water
at $T=20$~C \cite{Abdurakhimov:10}. The measurements of wave damping in
the cell showed that the quality factor at low frequencies $\omega <
\omega_d$ is $Q\sim 10^3$.

\noindent {\bf 2. Numerical Modeling.}  
The deviation of
the surface from the equilibrium flat state is expressed by time-dependent
amplitudes $a_k(t)$ of the normal modes \cite{Zakharov:67}. We assume
angular symmetry of the surface, so its deviation for capillary waves
is
$\zeta(r,t) = \sum_k \sqrt{ k / 2 \omega_k \rho A J_0(\beta_i)^2}$
$\times(a_k(t) + a_k^*(t))J_0(kr)$, 
where $r$ is the distance from the center of the cell, $J_0(x)$ is the
Bessel function of the zero order, $A$ is the free-surface area,
$\omega(k)$ is the linear dispersion relation given by Eq. \ref{eq:w},
$k \equiv
k_n = \beta_n /R$ is the radial wave number, $n >0$ is an integer
index labeling the resonant radial modes, and $\beta_n$ is the $n$th
zero of the first-order Bessel function $J_1(\beta_n)=0$.  In the
simulations, $r$ is measured in units of the capillary length scale $\lambda_c$,
and time $t$ is measured in the units of $t_c = \omega_c^{-1}$.  The
driving force is applied at a given radial mode $k_d$.  Due to angular
isotropy, we utilize the angle-averaged dynamical equation for
$a_k(t)$ \cite{Pushkarev:96},
\begin{eqnarray} 
  {d a_{k}(t) \over dt}  & = &   - i \sum_{k_1,k_2} V_{k, k_1,k_2}\,  D _{k, k_1, k_2} \,
    a_{k_1}(t)  a_{k_2}(t)   e ^{i(\omega(k) - \omega({k_1})  - \omega({k_2}))t}\nonumber \\
   &  -  &   2i  \sum_{k_1,k_2} V_{k_1, k, k_2}^* \, D _{k_1, k, k_2} \, a_{k_1}(t) a_{k_2}^* (t)   e ^{i(\omega(k) + \omega({k_2}) - \omega({k_1}))t}   
     -   \gamma(\omega({k})) a_{\bm{k}}(t).
    \label{eq:3w}
\end{eqnarray}
The coupling coefficients ${V}_{k, k_1, k_2}$ characterize the
interaction strengths between waves with wave numbers $k$, $k_1$, and
$k_2$; instead of taking the exact value for capillary waves, we model
it by ${V}_{k, k_1, k_2} = \epsilon \sqrt{\omega(k) \omega({k_1})
  \omega({k_2})}$ \cite{Zakharov:92}. Star denotes complex conjugate,
$i$ stands for the imaginary unit, and
$D_{k_1, k, k_2} = 1/2 \pi \Delta (k, k_1, k_2)$,
where $ \Delta (k, k_1, k_2) $ is the area of the triangle with sides
$k$, $k_1$, and $k_2$. We consider $n_{\rm max}=100$ radial modes.
The dimensionless factor $\epsilon \ll 1$ characterizes nonlinearity
of the system and is of the order of the maximum surface slope with
respect to the horizontal \cite{Zakharov:67}. We set $\epsilon
=10^{-2}$ as a representative value \cite{Brazhnikov:02a}. Due to the
small nonlinearity, we only retain three-wave interactions in
Eq.\ \ref{eq:3w}; the inclusion of four-wave scattering requires
special considerations \cite{During:09} and is deferred to future
studies.

Driving was at the 50th mode by fixing the wave amplitude $a_d \equiv
|a_{k_d}(t)|$ at a given value,  set in the present simulations as
$(\lambda_c^{7/2} \omega_c \rho)^{-1/2} a_d = 0.1$.  
We also
add wave damping at both high and low frequencies, to mimic the
physical effects that remove energy from the system. Specifically, we
model the wave damping coefficient as
\begin{equation}
 \gamma(\omega) = \gamma_{LF}(\omega) + \gamma_{HF}(\omega), \label{eq:Damping} 
\end{equation}
which is the sum of damping at low frequencies below the 10th
resonance in the cell, with $\gamma_{LF}(\omega) =
\gamma_{LF}g_{LF}(\omega)$, as well as damping at high frequencies
above the 80th resonance, with $\gamma_{HF}(\omega) =
\gamma_{HF}g_{HF}(\omega)$.  The range of wave frequencies between the
10th and 80th resonant frequencies can be considered as a ``numerical
inertial interval'' in which damping is absent.  The dimensionless
damping factor at high frequencies was set as $\gamma_{HF} = 5\times
10^{-2}\omega_c$.  Damping at high resonant numbers $n> n_{HF}=80$ is
modeled as $ g_{HF}(n) = (n - n_{HF})^2 / (n_{\rm max} - n_{HF})^2$,
and $ g_{HF}(n) = 0$ for $n \leq n_{HF}$.  For model waves on a fluid
layer of finite depth, we model damping at low resonant numbers
$n<n_{LF} = 10$ as $g_{LF}(n) = (n_{LF} - n) / n_{LF}$, and $g_{LF}(n)
=0$ for $n \geq n_{LF}$.  The low-frequency damping coefficient
$\gamma_{LF} $ is varied between $0$ and $2 \gamma_{HF}$.  To
calculate the dependence of $a_k(t)$ on time $t$, we integrated
Eq.\ \ref{eq:3w} until the system reached the steady state.  We
found numerical convergence and energy conservation with $10^{-7}$
numerical accuracy.  The wave spectrum is calculated as the
time-averaged quantity $N(k) = \langle |a_k(t)|^2 \rangle$.  For
capillary waves, the time-averaged correlation function is $I(\omega)
= N(k(\omega))$, where $N(k)$ is expressed as a function of the wave
frequency $\omega$ via the relation $k=k(\omega)$ from
Eq. \ref{eq:w} \cite{Brazhnikov:02a}.

\end{document}